\documentclass[%
 reprint,
 amsmath,amssymb,
 aps,
floatfix,
]{revtex4-2}

\usepackage{graphicx}
\usepackage{dcolumn}
\usepackage{bm}
\usepackage{hyperref}
\usepackage{booktabs}
\usepackage{gensymb}
\usepackage[utf8x]{inputenc}                          
\usepackage[english]{babel}
\usepackage{float}
\usepackage{placeins}
\usepackage{subcaption}
\usepackage{orcidlink}
\usepackage{ragged2e}
\usepackage{soul}
\usepackage{cleveref}
\crefname{figure}{Figure}{Figures}
\usepackage{caption}

\begin{document}

\preprint{APS/123-QED}

\title{Vector Measurements Using Integrated Radio Frequency Atomic Magnetometers}

\author{Ayse Marasli\,\orcidlink{0000-0003-4449-0243}}
\email{amarasli87@gmail.com}
\author{Karen L. Sauer\,\orcidlink{0000-0003-2784-8530}}%
 \email{ksauer1@gmu.edu}
\affiliation{ Quantum Science and Engineering Center, George Mason University, Fairfax, Virginia 22030, USA}
\author{Thomas W. Kornack\,\orcidlink{0000-0003-1615-5325}}
\email{kornack@twinleaf.com}
\author{D. Casey Oware}
 \email{oware@twinleaf.com}
\affiliation{Twinleaf LLC, 300 Deer Creek Drive Suite 300, Plainsboro, NJ 08536, USA}

\begin{abstract}
We demonstrate reconstruction of three-dimensional radio-frequency (RF) magnetic-field vectors using a pair of integrated RF atomic magnetometers operated with orthogonal bias-field orientations. A theoretical and experimental analysis identifies a phase-ambiguity dead band that limits reconstruction when the two sensor responses become nearly identical. Measurements performed in an unshielded laboratory environment demonstrate accurate reconstruction of RF magnetic-field orientations and validate the predicted dependence of reconstruction accuracy on signal imbalance. 
These results establish integrated RF atomic magnetometers as a compact and sensitive platform for directional RF magnetic-field sensing, particularly at low frequencies, and provide a foundation for portable source-localization and field-mapping applications.
\end{abstract}

\maketitle


\section{\label{sec:intro}Introduction\protect}

Vector measurements of radio-frequency (RF) magnetic fields provide information unavailable from scalar measurements alone, enabling applications such as field mapping and source localization of RF signals. These capabilities could be important in applications including magnetic resonance, electromagnetic compatibility testing, magnetic induction sensing, communications, and space physics. RF vector measurements are traditionally performed using tri-axial induction-coil sensors \cite{Tumanski2007}. Search coils offer direct vector measurements, broad bandwidth, and a mature technology, and are widely used in scientific and spacecraft instrumentation \cite{Fergeau2005,Lecontel2016}. However, their sensitivity decreases with frequency, often requiring larger coils or magnetic cores to maintain performance at low frequencies \cite{Tumanski2007}. In addition, vector accuracy depends on precise orthogonality and calibration of the sensing axes. Misalignment, gain and phase mismatches, cross-talk between coil elements, and  unwanted electric-field pickup, can introduce systematic errors that degrade field reconstruction \cite{Tong2025}. 

Atomic magnetometers are among the most sensitive magnetic sensors available, achieving femtotesla and even sub-femtotesla sensitivities without cryogenic cooling \cite{budker2007optical,allred2002high,kominis2003subfemtotesla}. Consequently, substantial effort has been devoted to the development of vector atomic magnetometers for close to DC magnetic fields, motivated by applications in navigation, geophysics, magnetic imaging, and biomagnetic sensing \cite{seltzer2004unshielded,patton2014all,lu2022triaxial,bulatowicz2023feedback,Liu2023,Meng2023MachineLA,Wang2025Pulsed}. For oscillating magnetic fields, a complementary approach is provided by radio-frequency atomic magnetometers (RF-AMs), which operate by tuning the atomic Larmor frequency into resonance with the signal of interest~\cite{savukov2005tunable,alexandrov2013mx}. Unlike induction-based sensors, RF-AMs provide near frequency-independent magnetic-field sensitivity across their tuning range while remaining intrinsically insensitive to direct RF electric-field pickup \cite{savukov2005tunable,savukov2007detection,Lee2006}. These characteristics, together with their compact sensing volume and negligible magnetic coupling between sensing elements, make RF-AMs attractive for portable sensors, magnetic-field mapping, and multi-sensor arrays, particularly at  low frequency~\cite{savukov2007detection}. Accordingly, RF-AMs have enabled sensitive detection of weak oscillating magnetic fields in applications including low-field nuclear magnetic resonance (NMR) \cite{savukov2005nmr,savukov2007detection}, nuclear quadrupole resonance (NQR) \cite{Lee2006,cooper2016atomic}, magnetic induction measurements and tomography \cite{Wickenbrock2014,Rushton2024}, and unshielded sensing \cite{keder2014unshielded,cooper2018jmr}.

Despite the high sensitivity of RF atomic magnetometers, most demonstrations have focused on measurement of a single component of an oscillating magnetic field. Recent advances have extended RF atomic magnetometry to polarization-sensitive measurements and determination of RF-field orientation within a transverse plane \cite{motamedi2023magnetic}.  In addition, vector reconstruction has been demonstrated using sequential measurements acquired at multiple sensor orientations when the temporal evolution of the field is known \cite{marasli2026}.  These developments have enabled characterization of RF-field polarization and localization of nearby magnetic dipole sources \cite{motamedi2023magnetic,marasli2026}. However, reconstruction of arbitrary three-dimensional RF magnetic-field vectors without sensor rotation or prior knowledge of the signal's timing remains an outstanding challenge.

In this work, we demonstrate reconstruction of three-dimensional RF magnetic-field vectors using simultaneous measurement of a pair of integrated RF atomic magnetometers operated with orthogonal bias-field orientations. Operating in an unshielded laboratory environment, the system reconstructs RF magnetic-field orientations without mechanical rotation.  Furthermore when the signal timing is not known a priori, reconstruction remains effective over much of the orientation space, with the associated limitations quantified by the dead-band analysis presented below. The demonstrated capability provides a compact platform for RF field mapping, source localization, and future determination of electromagnetic-wave propagation through combined electric and magnetic field measurements.

\section{\label{sec:theory}Theory \protect}

\subsection{Vector-field reconstruction}

Consider an RF magnetic field
\begin{equation}
\mathbf{B}(t)
=
\left(
B_x \hat{\mathbf{x}}
+
B_y \hat{\mathbf{y}}
+
B_z \hat{\mathbf{z}}
\right)
\cos(\omega t+\phi),
\end{equation}
where $\phi$ is a phase referenced to the detection system.  The present analysis is restricted to linearly polarized RF magnetic fields, for which the field orientation is described by the fixed vector $(B_x,B_y,B_z)$.  Two atomic sensors with orthogonal tuning fields, but a common probe direction along $\hat{\mathbf{x}}$, are used to reconstruct the vector field. Sensor~1 is tuned along $\hat{\mathbf{z}}$ and measures the transverse field in the $xy$ plane,
\begin{equation}
\mathbf{B}_1(t)
=
B_1
\cos(\omega t+\phi)
\left(
\cos\eta_1\,\hat{\mathbf{x}}
+
\sin\eta_1\,\hat{\mathbf{y}}
\right),
\end{equation}
while Sensor~2 is tuned along $-\hat{\mathbf{y}}$ and measures the transverse field in the $xz$ plane,
\begin{equation}
\mathbf{B}_2(t)
=
B_2
\cos(\omega t+\phi)
\left(
\cos\eta_2\,\hat{\mathbf{x}}
+
\sin\eta_2\,\hat{\mathbf{z}}
\right).
\end{equation}
The field components are therefore
\begin{align} \label{components}
B_x &= B_1\cos\eta_1 = B_2\cos\eta_2,\\
B_y &= B_1\sin\eta_1, \\
B_z &= B_2\sin\eta_2.
\end{align}
Hence, complete vector reconstruction will require knowledge of the common phase $\phi$.   In some cases, particularly ones in which a sample is excited, the phase is known, but in others the common phase must be determined.

We now relate these field components to the experimentally measured signals. The steady-state response to the applied orthogonal RF field is given by the atomic polarization along $\hat{\mathbf{x}}$~\cite{Alem2011SpinDamping},
\begin{equation}
P(t)
=
\frac{\gamma B_\perp T_2}
     {2\sqrt{1+\Delta\omega^2T_2^2}}
\cos\!\left(
\omega t+\phi+\eta
+\theta(\Delta\omega)-\frac{\pi}{2}
\right),
\end{equation}
where  $\gamma$ is the gyromagnetic ratio, $\Delta\omega=\omega-\omega_L$ and
$\theta(\Delta\omega)=-\tan^{-1}(\Delta\omega T_2)$.
After calibration to remove the resonance response and dispersive phase shift, the measured signals reduce to
\begin{align} \label{signals}
S_1(t) &= B_1\cos(\omega t+\phi+\eta_1), \\
S_2(t) &= B_2\cos(\omega t+\phi+\eta_2),
\end{align}
or in phasor notation
\begin{align} \label{phasors}
\tilde S_1 &= B_1 e^{i(\phi+\eta_1)} = e^{i\phi} (B_x + i B_y)\\
\tilde S_2 &= B_2 e^{i(\phi+\eta_2)} = e^{i\phi} (B_x + i B_z),
\end{align}
Therefore if $\phi$ is known, reconstruction is straightforward:
\begin{align}
B_x &= \frac12
      \mathrm{Re}
      \!\left[
      \lambda(\tilde S_1+\tilde S_2)
      \right],\\
B_y &= \mathrm{Im}
      \!\left[
      \lambda\tilde S_1
      \right],\\
B_z &= \mathrm{Im}
      \!\left[
      \lambda\tilde S_2
      \right],
\end{align}
where
\begin{eqnarray}
\lambda
=
e^{-i\phi}.
\end{eqnarray}

\begin{figure}[h]
   \centering
   \includegraphics[width=0.85\linewidth]{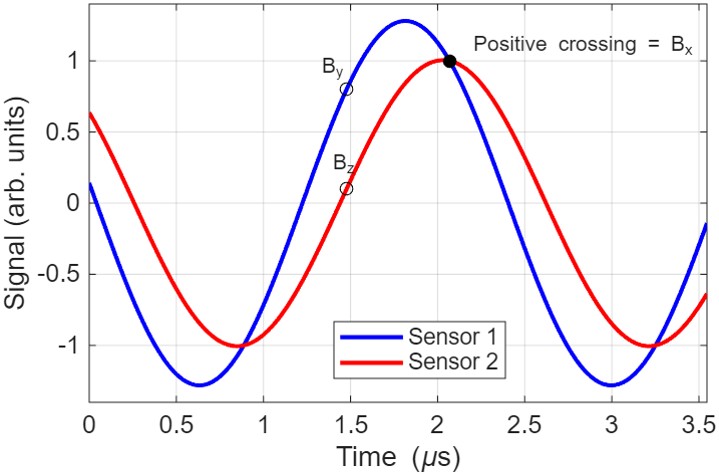} 
   \caption{\justifying Simulated calibrated responses of the two orthogonal RF sensors for an arbitrary RF-field orientation. The positive signal crossing occurs at $t=t_0$, where both sensors measure the common component $B_x$. One quarter cycle earlier, the sensor outputs correspond to $B_y$ and $B_z$.}
\label{fig:2sensor}
\end{figure}

One way of finding $\phi$ is to notice that there exists a time $t_0$ when the two signals become equal 
\begin{equation}
S_1(t_0)=S_2(t_0)=B_x,
\end{equation}
corresponding to
\begin{equation}
\omega t_0=-\phi,
\end{equation}
as can be seen through Eqs.~\ref{components} and \ref{signals}.
Note that this condition is generally met twice per cycle, corresponding to an arbitrary choice of sign for $B_x$.
A quarter cycle earlier,
\begin{align}
S_1\!\left(t_0-\frac{\pi}{2\omega}\right) &= B_y,\\
S_2\!\left(t_0-\frac{\pi}{2\omega}\right) &= B_z.
\end{align}
Figure~\ref{fig:2sensor} illustrates this principle.  Zero-crossing of the signal difference could therefore be utilized to determine $\phi$.

While the above time-domain discussion provides an intuitive view of the phase determination, the use of phasors is more convenient for analysis and directly matches data from a phase-sensitive spectrometer. 
From Eqs.~\ref{phasors}, constrained by Eq.~\ref{components}, we obtain
\begin{equation}  \label{lambda}
\lambda =
\pm i\,
\frac{|\tilde S_1-\tilde S_2|}
     {\tilde S_1-\tilde S_2},
\end{equation}
where the $\pm$ sign reflects the fact that the spatial orientation of a time-varying RF magnetic field is defined only up to an overall sign. 
It is clear from Eq.~\ref{lambda}, the reconstruction becomes singular when
\begin{equation}
\tilde S_1=\tilde S_2,
\end{equation}
since the phase factor $\lambda$ is then undefined. This condition corresponds to $B_y=B_z$ and produces a dead-band in field orientation space. The impact of measurement noise on this singularity was evaluated using Monte Carlo simulations (1,000 iterations) with fixed relative uncertainty $\sigma_B/B$.

\subsection{Dead-band analysis for unknown phase}

\begin{figure}
   \centering
   \includegraphics[width=\linewidth]{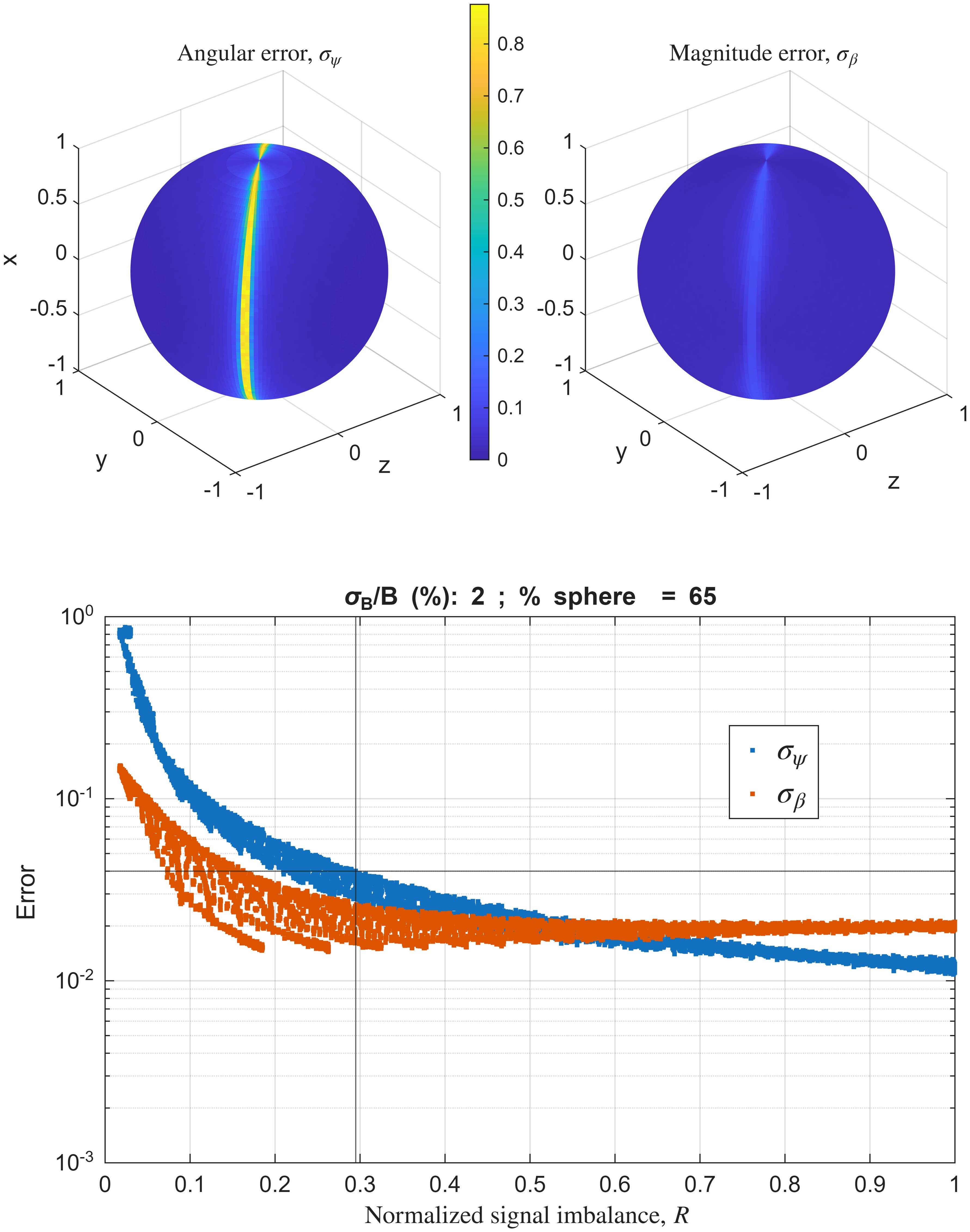} 
\caption{\justifying Monte Carlo simulation of reconstruction accuracy
for $\sigma_B/B=2\%$.
Top: spherical maps of the angular error
$\sigma_\psi$ and relative amplitude error
$\sigma_\beta$.
Bottom: Errors as a function of the normalized signal imbalance
$R$.
The vertical line denotes the threshold value $R_{\rm th}$, the maximum value for which $ \sigma_\psi > 2\sigma_B/B$, the latter denoted by a horizontal line.  The two lines break the graph into quadrants.  The lower right quadrant is a region of accurate vector reconstruction, and the corresponding angular coverage is given in the title. }
\label{fig:HeadingError}
\end{figure}

To quantify reconstruction accuracy, the recovered field
$\mathbf{B}'$ was compared with the applied field
$\mathbf{B}$. Angular and amplitude deviations were defined as
\begin{align}
\psi  &= \cos^{-1}\!\left(\hat{\mathbf n}\cdot\hat{\mathbf n}'\right), \\
\beta &= 1 - \frac{B'}{B}.
\end{align}
where $\hat{\mathbf n}$ and $\hat{\mathbf n}'$ are the corresponding unit vectors. 
 Figure~\ref{fig:HeadingError} shows the standard deviation of $\psi$ and
$\beta$ for $\sigma_B/B=2\%$. The largest errors form a narrow ring
corresponding to the singular condition $B_y=B_z$, where the two sensor
responses become nearly identical. Away from this region the error
rapidly approaches the measurement-noise limit.

The proximity to the singularity is characterized by the normalized
signal imbalance
\begin{equation}
R= \frac{|\tilde{S}_1-\tilde{S}_2|}{|\tilde{S}_1|+|\tilde{S}_2|}.
\end{equation}
Both $\sigma_\psi$ and $\sigma_\beta$ increase sharply as
$R\rightarrow0$, reflecting the loss of phase sensitivity near the
dead-band. A practical operating region can be defined by
\begin{equation}
\sigma_\psi < 2\frac{\sigma_B}{B},
\end{equation}
which yields a threshold value $R_{\rm th} \approx 0.3$ for
$\sigma_B/B=2\%$.  As shown in Fig.~\ref{fig:HeadingError}, these two thresholds are used to define a region in which both $\sigma_\psi$ and $\sigma_\beta$ on the order of the normalized field error. 

\begin{figure}
   \centering
   \includegraphics[width=\linewidth]{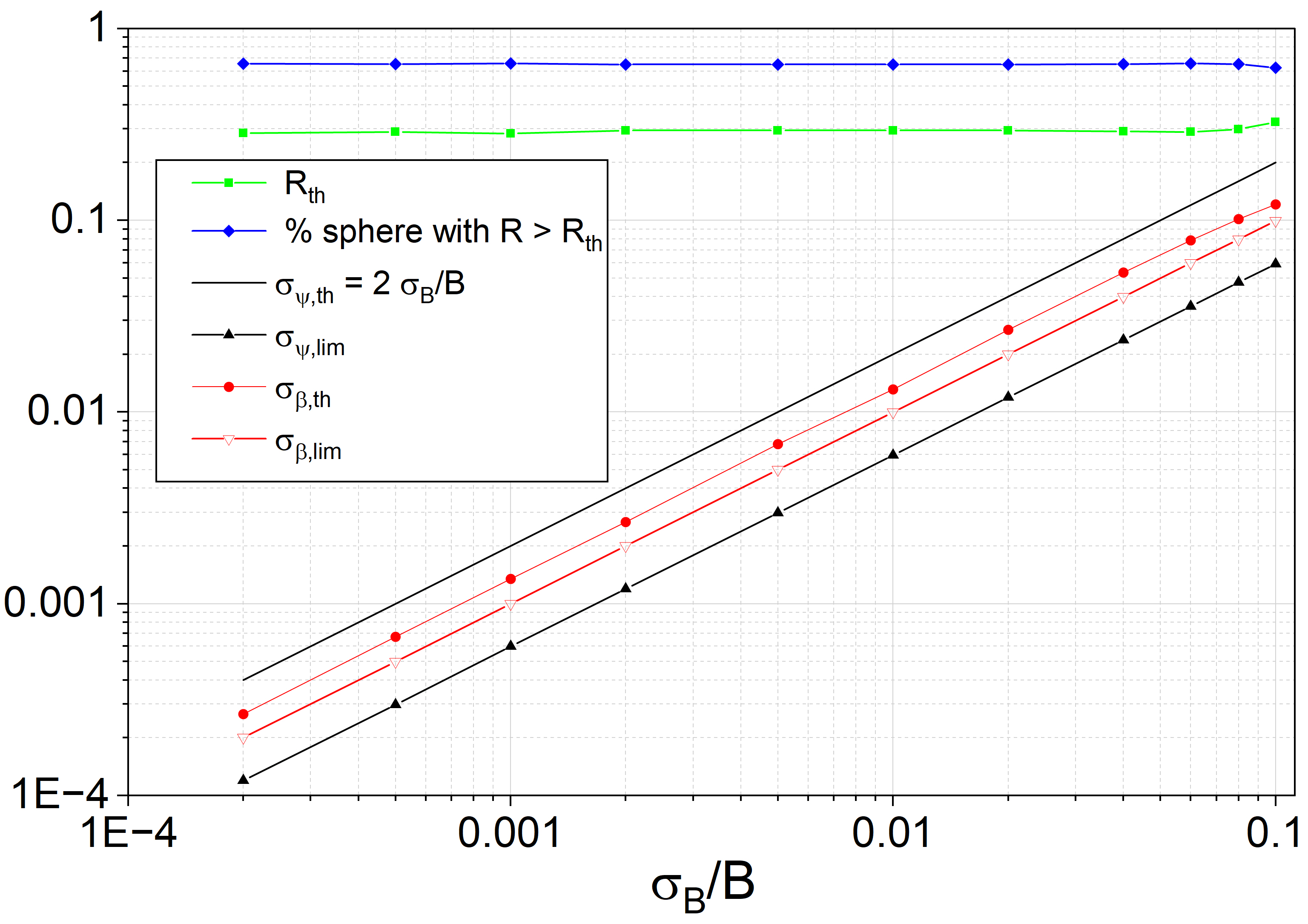} 
   \caption{\justifying Dependence of reconstruction performance on the
relative measurement uncertainty $\sigma_B/B$. The asymptotic
angular and amplitude errors,
$\sigma_{\psi,\mathrm{lim}}$ and
$\sigma_{\beta,\mathrm{lim}}$, as well at the value of $\sigma_{\beta}$ at $R_{\rm th}$,
scale with measurement uncertainty, whereas the threshold imbalance
$R_{\rm th}$ and the corresponding angular coverage vary weakly with
measurement noise. }
\label{fig:SummaryHE}
\end{figure}

Figure~\ref{fig:SummaryHE} summarizes the dependence on measurement
uncertainty.   The asymptotic reconstruction errors obtained far from the dead-band
($R\rightarrow1$) track with $\sigma_B/B$.  In contrast, $R_{\rm th}$ and the corresponding angular
coverage exhibit only weak dependence on noise level. 
For small
$\sigma_B/B$, the threshold approaches
$R_{\rm th}\approx 0.3$, corresponding to an angular coverage of
approximately $65\%$ of the sphere.

The normalized imbalance $R$ therefore serves as a practical indicator
of reconstruction fidelity in the case where $\phi$ is unknown. Because it can be computed directly from the
measured sensor responses, it provides a simple criterion for assessing
whether a given field orientation lies sufficiently far from the
dead-band to permit reliable vector reconstruction.  

\section{\label{sec:Experimental}Experimental Setup\protect}

\subsection{\label{sec:level2}Characterization of Integrated RF Atomic Magnetometer\protect}

\begin{figure}
\subfloat{\includegraphics[width=0.9\linewidth]{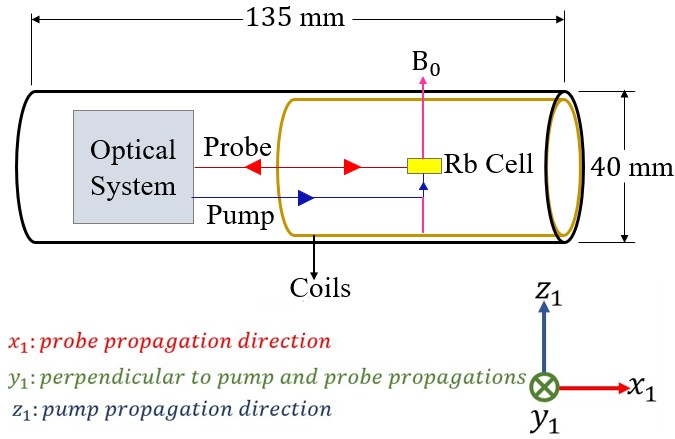}}
\caption{\justifying Schematic of Sensor~1 viewed along the $y_1$ axis. The atomic vapor cell (yellow), crossed pump (blue) and probe (red) laser beams, and field coils (brown) that generate the bias magnetic field $\mathbf{B}_0$ (magenta) are enclosed within a cylindrical housing.}
\label{fig:sensor_head}
\end{figure}

Each of the two magnetometers in the integrated sensor head is based on a vapor cell with dimensions of $3.5 \times 3.5 \times 9\,\mathrm{mm}$, with the long axis aligned along the probe beam direction. The cell is filled with $^{87}\mathrm{Rb}$ atoms, $0.6$ amagat of neon buffer gas, and $0.1$ amagat of nitrogen as a quenching gas. The probe beam traverses the cell twice. The cell is housed in a resistive oven that maintains a saturated vapor pressure corresponding to an atomic number density of approximately $2 \times 10^{13}\,\mathrm{atoms/cm^3}$ in Sensor 1, while the number density in Sensor 2 was approximately half that value. The optical arrangement consists of a pump laser tuned to the D1 transition and an off-resonant probe laser detuned by a few tenths of a nanometer from the D1 resonance. The beams are shaped and polarized using collimating optics, and the probe polarization rotation is detected by a balanced polarimeter equipped with photodiodes ~\cite{chauvat1997magnification,makarov2023observation}. Surrounding the optics and the cell are three orthogonal coil pairs, which provide a bias magnetic field and determine the resonance frequency. All components are integrated into a cylindrical housing with a length of 13.5 cm and a diameter of 4 cm, as shown in Fig.~\ref{fig:sensor_head} for one magnetometer. The overall system includes the sensor head, a differential amplifier for the polarimetric signal, and a bench-top current driver controlled through LabVIEW software. For simplicity and reliability, signal transmission between modules is handled with electrical cables rather than optical fibers.

The performance of the system was characterized after optimization of the laser operating conditions. Under these conditions, the spin-spin relaxation time $T_2$, obtained from free-induction-decay measurements, was found to be $0.28~\mathrm{ms}$ for Sensor~1 and $0.24~\mathrm{ms}$ for Sensor~2.

\begin{figure}
\subfloat{\includegraphics[width=0.9\linewidth]{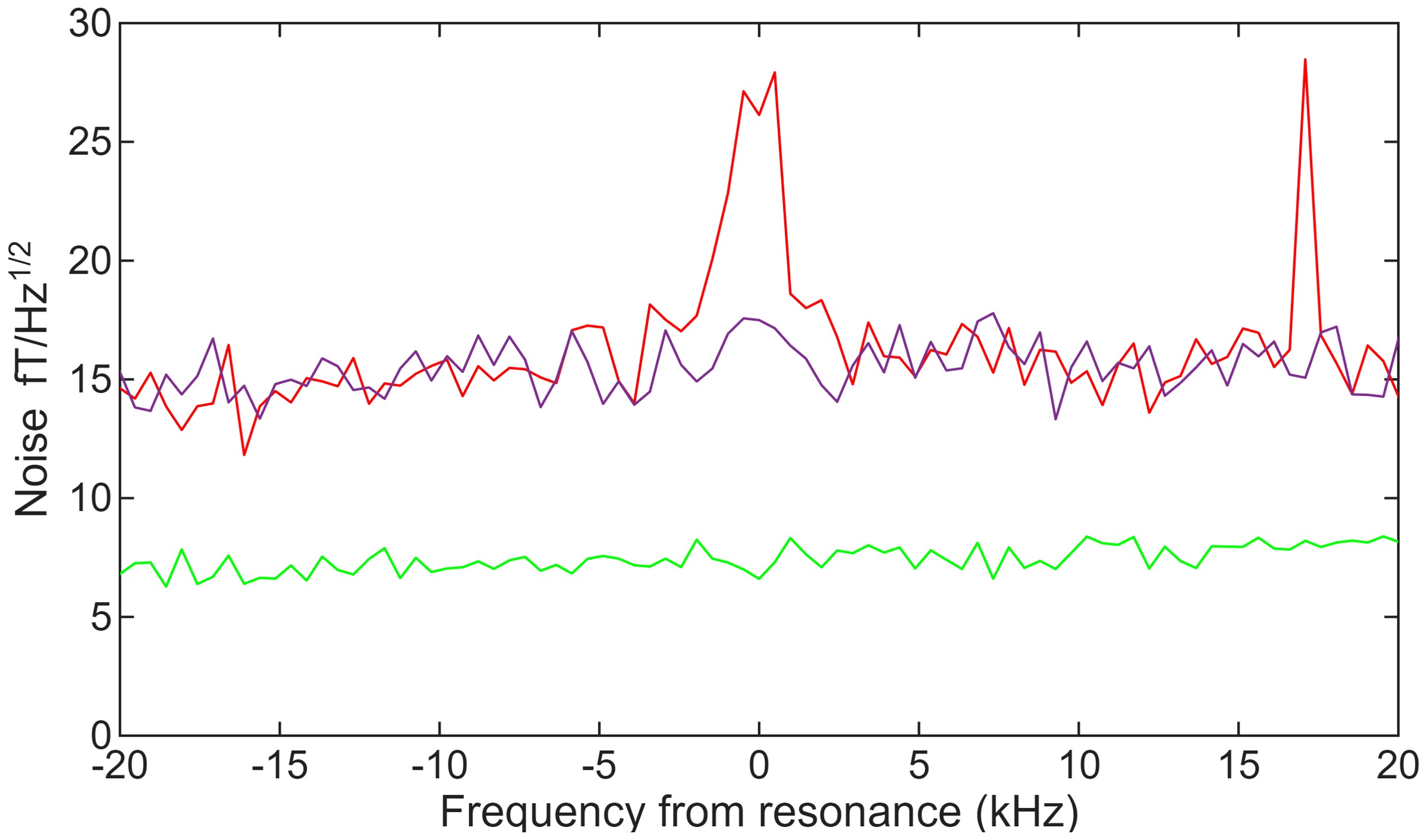}}
\caption{\justifying Sensor~1 sensitivity measured under RF-shielded
conditions, with the pump beam on (red), with the pump beam off (purple), and with the probe beam off (green).}
\label{fig:fig5}
\end{figure}

Different noise sources were identified by selectively enabling and disabling the pump and probe beams. Magnetic noise appeared only when the pump beam was active and was concentrated near the Larmor resonance frequency. Photon shot noise exhibited a white spectrum and was observed only when the probe beam was present. Electronic noise originating from the balanced polarimeter was present when the photodiodes were powered, while technical noise was found to be negligible when the photodiodes were switched off.
The sensitivity measurements were performed at a resonance frequency of $380~\mathrm{kHz}$. As shown in Fig.~\ref{fig:fig5}, Sensor~1 exhibited a resonant noise peak of $27~\mathrm{fT}/\sqrt{\mathrm{Hz}}$, corresponding to a factor of about 1.7 above the background noise level. The measurement was performed under RF-shielded conditions, indicating that the observed magnetic noise was predominantly associated with the sensor electronics. When operated in an unshielded laboratory environment, the resonant noise increased by 15\%.
In contrast, Sensor~2 exhibited a flat noise spectrum with a magnitude of $56~\mathrm{fT}/\sqrt{\mathrm{Hz}}$, consistent with a noise floor dominated by photon shot noise.

\subsection{Magnetic Field Vector Measurement Setup\protect}

\begin{figure}[h!]
\includegraphics[width=\linewidth]{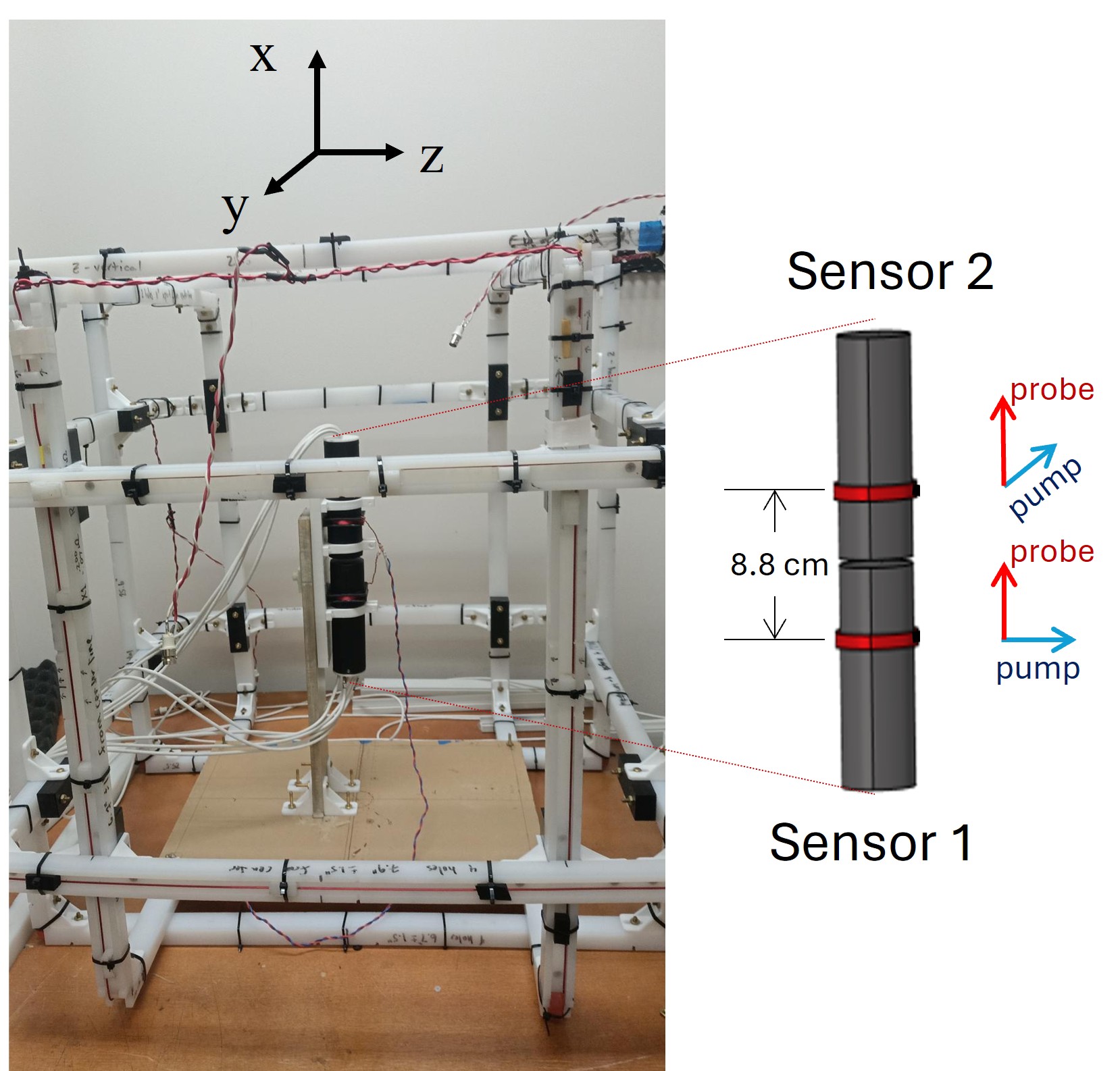}
\caption{\justifying The integrated sensor head is centered within a nested trio of square Helmholtz coils used to generate test magnetic field vectors. The callout of the dual magnetometers shows the orthogonal pump beams, as well as the placement of the calibration coils.}
\label{fig:fig6} 
\end{figure}

To generate homogeneous magnetic fields along three orthogonal directions on the surface of a sphere, three pairs of Helmholtz coils were constructed, each with a side length of 0.6~m, as shown in Fig.~\ref{fig:fig6}. To avoid self-resonance effects, each coil was wound with a single turn, and plastic frames were used to prevent magnetic shielding and eddy-current effects. 

An RF magnetic field at 423~kHz, corresponding to the NQR frequency of ammonium nitrate, was generated by 
controlling the currents to the Helmholtz pairs using the transmitter channels of a Tecmag Redstone spectrometer. The experiment was synchronized with the AC power line to suppress phase jitter.  In the composite sensor, centered within the Helmholtz coils, the bottom sensor is sensitive to the $x$ and $y$ components of the magnetic field ($B_x + iB_y$), while the top sensor is sensitive to the $x$ and $z$ components ($B_x + iB_z$).

\begin{figure}
\includegraphics[width=0.85\linewidth]{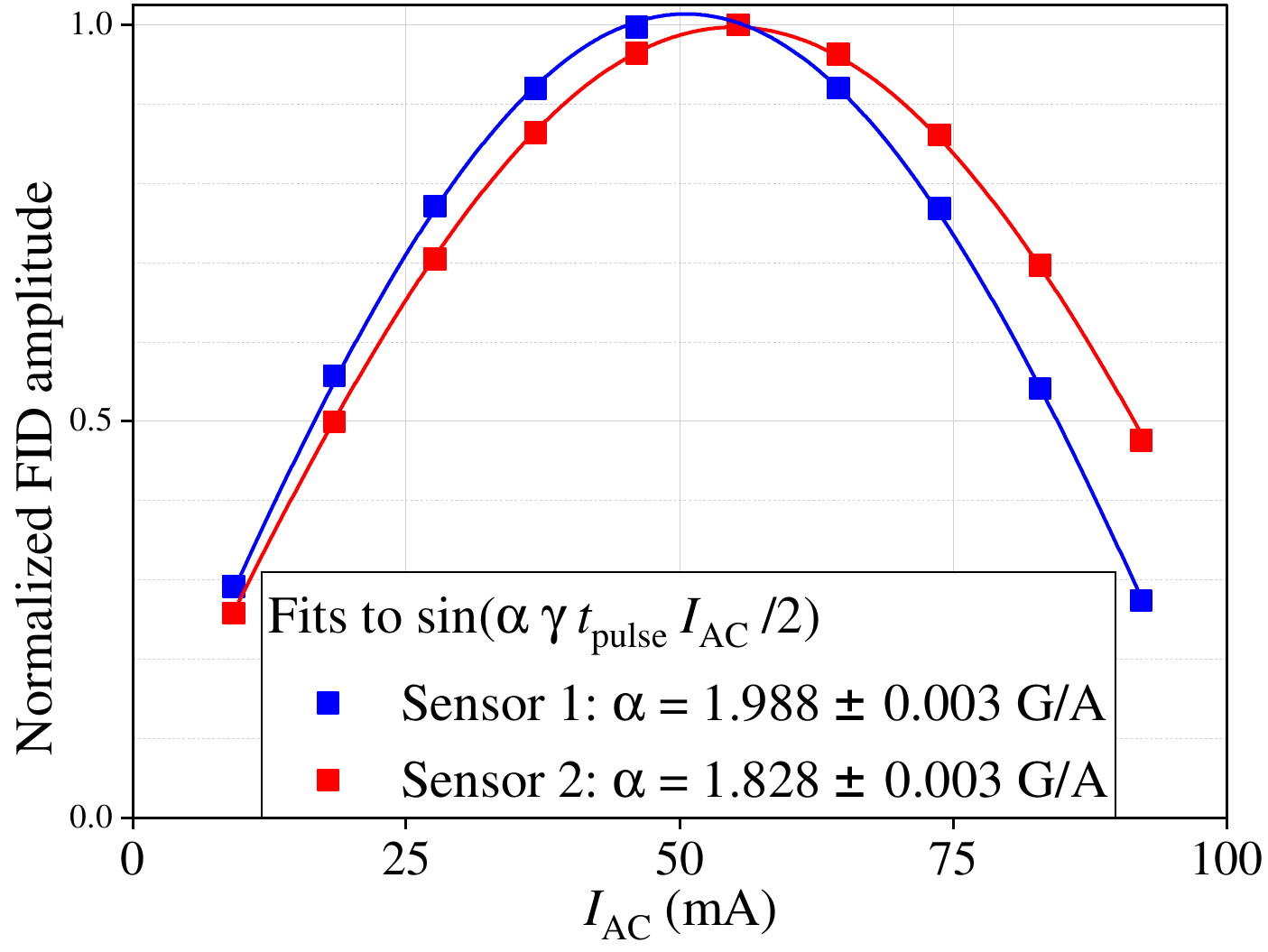}
\caption{\justifying Experimental determination of the magnetic field generated by the calibration coils is set by the calibration factor $\alpha$. A pulse length of $t_{pulse}$ = 7.1~$\mu$s was used to produce the FID signal at 423~kHz.}
\label{fig:fig7}
\end{figure}

In addition, to suppress the effects of environmental magnetic fluctuations, compensate for unequal sensor sensitivities, and determine the absolute magnetic-field values, calibration coils consisting of five turns each were connected in series. They were fixed such that each coil was centered on an alkali-atom vapor cell, as shown in Fig.~\ref{fig:fig6}.  Using a free-induction-decay (FID) after a short pulse, the magnetic field produced by the coil could be determined, as shown in Fig.~\ref{fig:fig7}.

\begin{figure}
   \centering
   \includegraphics[width=0.9\linewidth]{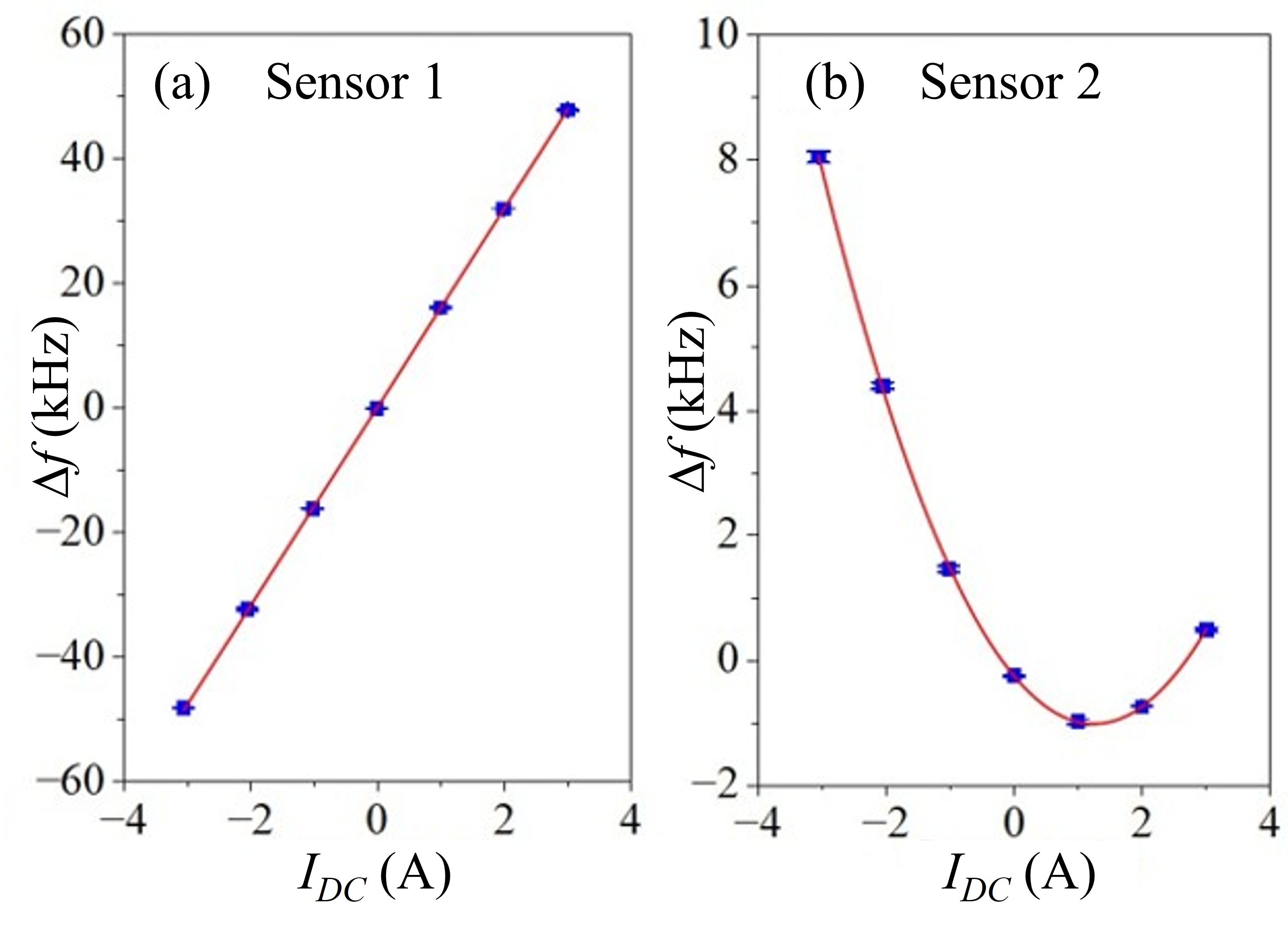} 
   \caption{\justifying Representative experimental characterization of the magnetic field produced by a Helmholtz pair using the frequency shift of the Larmor frequency.  Because the field variation in this case is along $z$, the shift in Sensor 1 is linear and the shift in Sensor 2 is quadratic.}
\label{fig:fig8}
\end{figure}

The Helmholtz pairs were 100 times weaker than the calibration coil. Direct calibration of the RF field via resonant spin excitation was limited by the long pulse durations required relative to $T_2$. Therefore, the field strength was characterized from shifts of the Larmor frequency produced by DC currents $I_{DC}$ in the coils. Representative data are shown in Fig.~\ref{fig:fig8} for the Helmholtz pair that produces a magnetic field along $z$, which is parallel to the tuning field of Sensor~1 and orthogonal to that of Sensor~2.  Therefore for Sensor~1 the frequency shift 
is $\Delta f = \gamma \alpha I_{DC} /(2 \pi)$, with $\alpha = 22.8 \pm 0.1~\mathrm{mG/A}$. In contrast, the frequency shift for Sensor~2 is 
\begin{eqnarray}
\Delta f  & = & \sqrt{ f_0^2 + \left( \frac{\gamma}{2 \pi} \alpha I_{DC} \right)^2} - f_0 \\
& \approx &  \frac{1}{2 f_0}\left(\frac{\gamma}{2 \pi} \alpha I_{DC} \right)^2  
\end{eqnarray}
where $f_0$, the unperturbed Larmor frequency, was 260~kHz. The resulting value of $\alpha$ for Sensor~2,  $22.7 \pm 0.1~\mathrm{mG/A}$,  is in agreement, within error bars, with the value for Sensor~1.

\begin{figure} 
\includegraphics[width=0.95\linewidth]{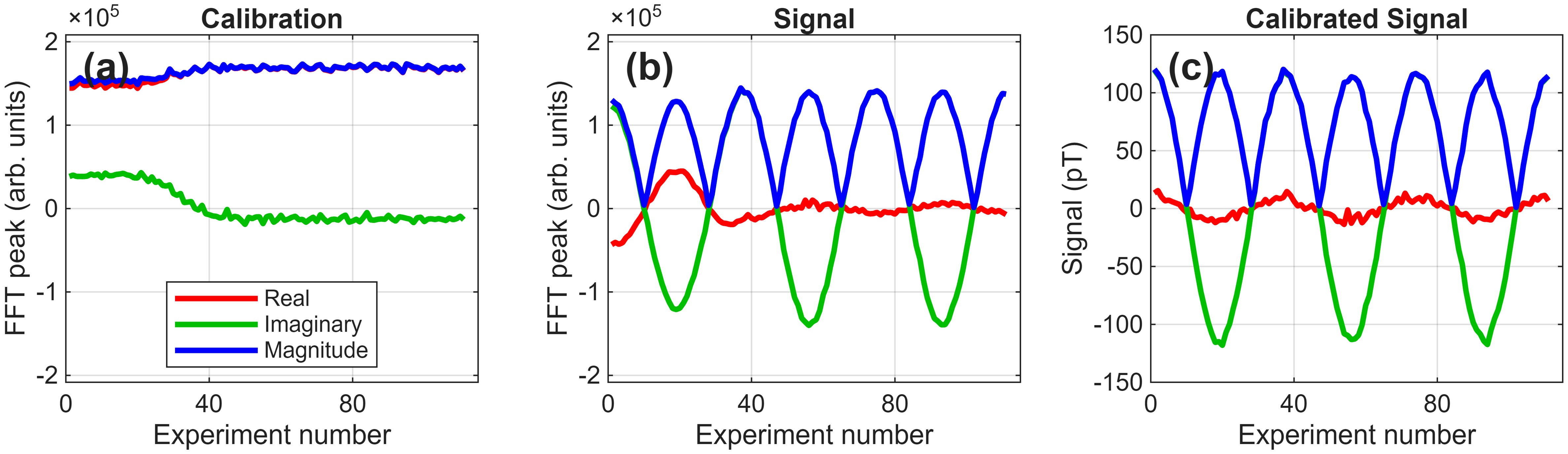}
\caption{\justifying Calibration and magnetic-field reconstruction for Sensor 2 during rotating-field excitation (at 423 kHz) in the \(y/z\) plane.
(a) Calibration response measured using a reference field applied along $x$. 
(b) Signal measured during excitation with a constant-magnitude RF field, whose direction is rotated in the \(y/z\) plane. 
In (a-b), the plotted quantities are the real, imaginary, and magnitude components of the FFT peak extracted from a 2-ms acquisition window.
(c) Reconstructed magnetic-field components after application of the calibration correction.}
\label{fig:fig9}
\end{figure}

\FloatBarrier

\section{\label{sec:result}Results and Discussion\protect}

\subsection{Vector-field reconstruction}

Radio-frequency magnetic field components in the \(x/y\), \(x/z\), and \(y/z\) planes were measured using the composite sensor with real-time in situ calibration. As an example, Fig.~\ref{fig:fig9} illustrates the calibration and signal-reconstruction procedure for Sensor~2 during excitation of the Helmholtz coils in the \(y/z\) plane. The applied field vector was rotated within the plane such that its tip traced a circular path, with measurements acquired at \(10^\circ\) angular increments. A calibration measurement was performed between successive vector measurements. Figure~\ref{fig:fig9}(a) shows the raw response of Sensor~2 to a reference calibration field applied along the \(x\)-axis. Slow temporal variations in the calibration signal are evident, likely reflecting environmental fluctuations occurring during the measurement sequence (\(<1\) min). Figure~\ref{fig:fig9}(b) shows the raw response during the rotating-field excitation, where a similar temporal phase variation is observed. Using the calibration measurements, the sensor responses were converted to magnetic-field units, yielding the reconstructed field components shown in Fig.~\ref{fig:fig9}(c). For this data set, a common phase adjustment, \(\lambda_0\), was applied to all measurements. The value of \(\lambda_0\) was determined as the weighted average of all measured \(\lambda\) values with normalized imbalance \(R>0.3\), using \(R\) as the weighting factor.

\begin{figure*}[t] 
    \centering
    \includegraphics[width= 0.95\textwidth]{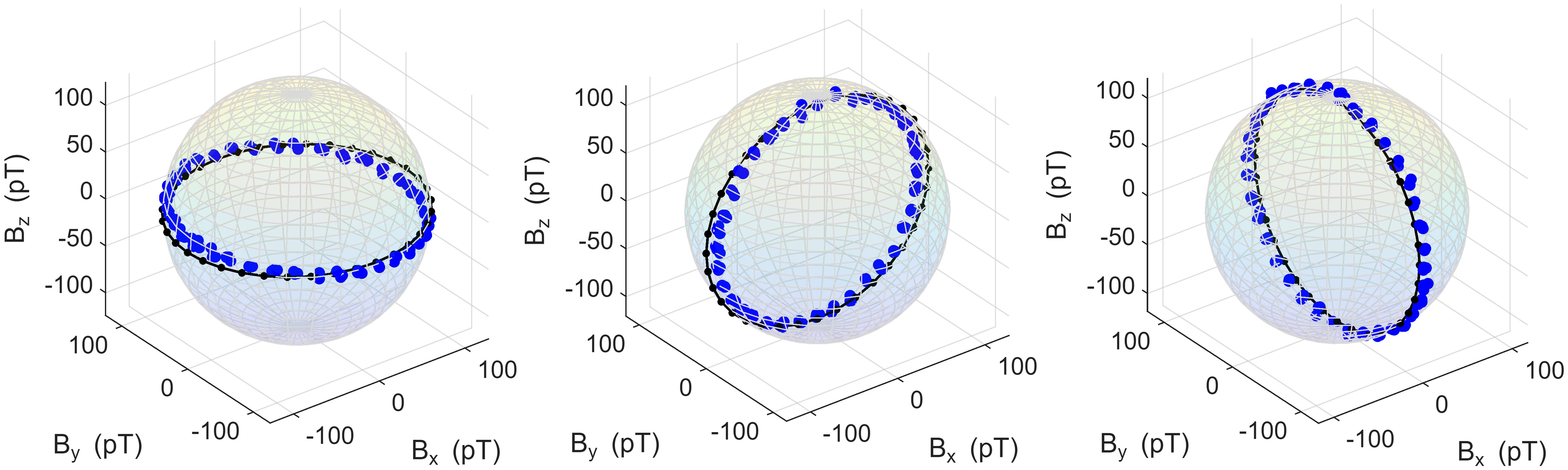}
    \caption{\justifying Magnetic field vectors generated in (a) the $x/y$ plane (b) the $x/z$ plane, and (c) the $y/z$ plane.  Blue circles represent measured data with a fixed common phase factor $\lambda_0$.  Black circles are predicted fields based on current values in the Helmholtz pairs.}
    \label{fig:fig10}
\end{figure*}

\begin{figure}
\centering
\includegraphics[width=0.95\linewidth]{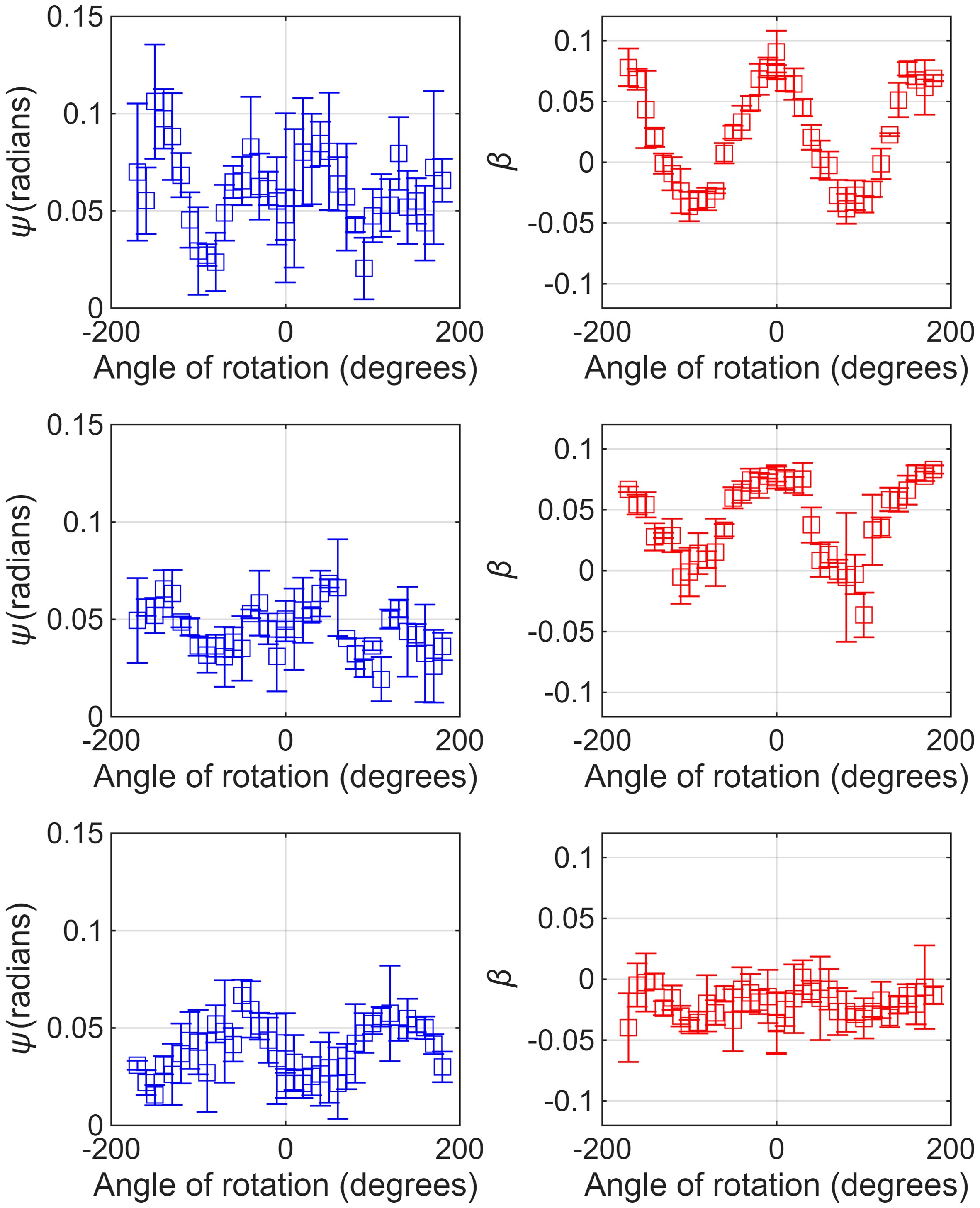}
\caption{\justifying
Angular deviation $\psi$ (left) and magnitude deviation $\beta$ (right) for the data in Fig.~\ref{fig:fig10} as a function of vector rotation angle.
Periodic patterns are indicative of systematic errors associated with the setup.
Top: $x/y$ (rotation angle from $x$); middle: $x/z$ (rotation angle from $x$); bottom: $y/z$ (rotation angle from $y$).}
\label{fig:fig11}
\end{figure}

The same procedure was repeated for rotations in each of the three orthogonal planes, and the resulting reconstructed field vectors are summarized in Fig.~\ref{fig:fig10}. In all cases, the measured vectors show good overall agreement with the vectors predicted from the coil-current model, although small but repeatable discrepancies are apparent. These residual errors are more clearly revealed in Fig.~\ref{fig:fig11}, which shows the angular deviation, \(\psi\), and fractional magnitude deviation, \(\beta\), as functions of rotation angle. Depending on the plane of rotation, either the angular or magnitude residual exhibits a stronger periodic structure. 

The periodic nature of these residuals suggests that the dominant error sources are systematic rather than random. Angular variations were on the order of few degrees.  Possible contributors include misalignment of the composite sensor relative to the Helmholtz coil assembly and deviations of the coil axes from ideal orthogonality.  The largest magnitude deviations, approximately $9\%$, occur when the RF field is directed along $x$, as shown in both Fig.~\ref{fig:fig10} and Fig.~\ref{fig:fig11}. As a result, the agreement between measurement and prediction is better in the $y/z$ plane, where the field has no $x$ component, than in the $x/y$ and $x/z$ planes.
  
This discrepancy corresponds to the measured field being smaller than predicted.  Most likely, this variation was due to uncertainty in the
test-coil calibration.  In particular, the field calibrations were performed using DC currents, whereas the vector measurements were acquired at 423~kHz.  Frequency-dependent effects, including complex impedances and parasitic capacitive or inductive coupling could therefore alter the effective field generation and contribute to the observed deviations. Notably, the calibration coil is aligned along the $x$ direction and exhibits a current sensitivity that is two orders of magnitude higher than that of the large Helmholtz coil used to generate the $B_x$ test field. As a result, parasitic inductive coupling—particularly involving the calibration coil—represents a plausible source of the observed discrepancy in the $B_x$ measurements.

\subsection{Dead-band analysis for unknown phase}

\begin{figure} 
    \centering
    \includegraphics[width=
    \linewidth]{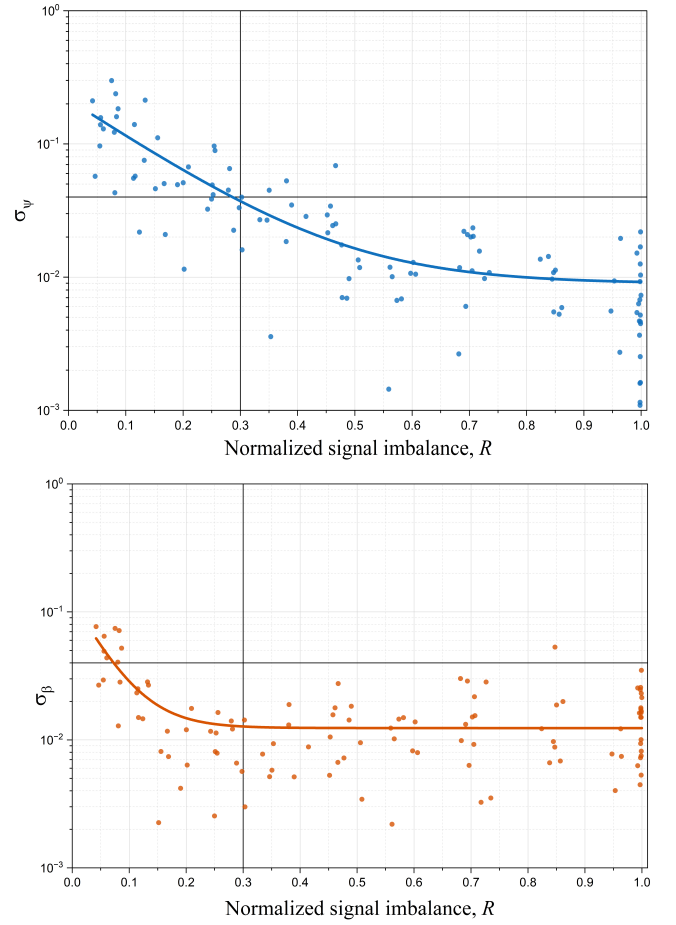}
    \caption{\justifying  Combined data for magnetic field vectors generated in all three planes, the error in $\psi$ and $\beta$ as function of the normalized difference.  Solid lines are fits of the data to an exponential function with an offset and are only meant to guide the eye.}
    \label{fig:fig12}
\end{figure}
\FloatBarrier

In the above section a fixed common phase factor $\lambda_0$ was used.  In this section we explore the experimental relationship between errors in $\psi$ and $\beta$ as a function of the normalized signal imbalance, $R$, with the phase factor calculated separately for each experiment.  In this case, only the statistical error bar is included, defined as the standard deviation of repeated measurements for the same vector orientation, in order to avoid the inclusion of systematic effects. The plots are shown in Fig.~\ref{fig:fig12}.  The plots are consistent with the predicted response to $R$, namely $R$ approaching 1 corresponds to much smaller deviations, and $R<0.3$ corresponds to significantly larger deviations. As shown in Figs.~\ref{fig:SummaryHE} and \ref{fig:HeadingError}, for $R\rightarrow1$ the asymptotic errors approach $\sigma_\beta \approx \sigma_B/B$ and $\sigma_\psi \approx (\sigma_B/B)/2$. Therefore the observed asymptotic errors are consistent with an effective normalized field uncertainty of approximately 2\%. In Fig.~\ref{fig:fig12}, the horizontal line corresponds to $R=0.3$, while the vertical line denotes an error of $4\%$, consistent with the criterion used in Figs.~\ref{fig:SummaryHE} and \ref{fig:HeadingError} to define the threshold imbalance.

Based on the sensitivity of the magnetometers, the reliance on calibration, and the 2 ms acquisition window, the predicted relative error would be on the order of 1\%.  This estimate does not include fluctuations in the power supplied to the Helmholtz coils. The experimentally observed errors are therefore broadly consistent with the expected level of uncertainty. Moreover, the data demonstrate that the normalized signal imbalance, $R$, provides a practical metric for predicting vector-reconstruction accuracy.

\FloatBarrier

\section{\label{sec:conclusion}Conclusion \protect}

We have demonstrated vector reconstruction of RF magnetic fields using a pair of integrated atomic magnetometers operated with orthogonal bias-field orientations in an unshielded environment. Measurements of RF magnetic fields whose orientations were systematically varied within three orthogonal planes verified recovery of the field orientation without mechanical sensor rotation. The remaining deviations were attributable primarily to field-generation and calibration uncertainties rather than limitations of the reconstruction formalism.

A key result of this work is the characterization of vector reconstruction in the absence of a known common phase. Theoretical analysis identified a dead-band associated with nearly identical sensor responses, and Monte Carlo simulations predicted increased angular and magnitude uncertainty as this condition is approached. Experimental measurements were consistent with these predictions and demonstrated that the normalized signal imbalance, $R$, provides a practical metric for estimating reconstruction accuracy. For the majority of the directional space, the error in angle and amplitude is on the same order as the field measurement error, corresponding to $R \gtrsim 0.3$.

These results establish integrated RF atomic magnetometers as a practical platform for directional RF magnetic-field sensing and provide an experimentally accessible metric for assessing vector-reconstruction accuracy in real time. The demonstrated capability is applicable to low-field NMR, NQR, magnetic-source localization, and vector mapping of RF magnetic fields. In combination with electric-field measurements, the approach may also enable determination of electromagnetic wave propagation direction through reconstruction of the Poynting vector.

\section{\label{sec:Acknowledgment}Acknowledgment\protect }
Support for this project comes from the Office of the
Undersecretary of Defense, Director of Defense Research
and Engineering for Modernization (DDRE(M)) Quantum
Science Office Contract 47QFLA23C0002.

\bibliography{Bio_Vector} 
\end{document}